\documentclass[a4paper,11pt]{article}  
\usepackage{aaskaiid}
\usepackage{orcidlink}
\usepackage{floatrow}

\title{Unique Science Opportunities for Space VLBI Systems with the SKA Telescopes}
\ShortTitle{Space VLBI with SKA}

\author[1]{Y.~Y.~Kovalev\orcidlink{0000-0001-9303-3263}}
\ShortName{Kovalev et al.} 
\author[2]{G.~Bruni\orcidlink{0000-0002-5182-6289}}
\author[3]{T.~An\orcidlink{0000-0003-4341-0029}}
\author[4,5]{H.~E.~Bignall\orcidlink{0000-0001-6247-3071}}
\author[6]{P.~G.~Edwards\orcidlink{0000-0002-8186-4753}}
\author[7]{C.~Garc\'ia-Mir\'o\orcidlink{0000-0003-1136-5016}}
\author[8]{M.~Giroletti\orcidlink{0000-0002-8657-8852}}
\author[9,10]{L.~I.~Gurvits\orcidlink{0000-0002-0694-2459}}
\author[11]{M.~Kadler\orcidlink{0000-0001-5606-6154}}
\author[12]{J.-Y.~Kim}
\author[13]{E.~V.~Kravchenko\orcidlink{0000-0003-4540-4095}}
\author[14,1]{I.~Liodakis\orcidlink{0000-0001-9200-4006}}
\author[15]{Y.~Q. Liu\orcidlink{0000-0001-9321-6000}}
\author[16]{A.~V.~Plavin\orcidlink{https://orcid.org/0000-0003-2914-8554}}
\author[17,13]{A.~V.~Popkov}
\author[18,13]{A.~B.~Pushkarev\orcidlink{0000-0002-9702-2307}}
\author[19,20,1]{T.~Savolainen}
\author[15,21]{Z.-Q. Shen\orcidlink{0000-0003-3540-8746}}
\author[1] {A.~Tamar\orcidlink{0000-0001-8763-4169}}
\author[22,1]{E.~Traianou\orcidlink{0000-0002-1209-6500}}

\affiliation[1]{Max Planck Institute for Radio Astronomy, Auf dem Huegel 69, 53121 Bonn, Germany}
\emailAdd{yykovalev@gmail.com}
\emailAdd{adityatamar@gmail.com}
\affiliation[2]{Institute for Space Astrophysics and Planetology, INAF, via del Fosso del Cavaliere 100, I-00133 Rome, Italy}
\emailAdd{gabriele.bruni@inaf.it}
\affiliation[3]{Department of Astronomy, University of Science and Technology of China, 96 Jinzhai Road, Hefei, Anhui 230026, China}
\emailAdd{antao2008@ustc.edu.cn}
\affiliation[4]{Manly Astrophysics, 15/41-42 East Esplanade, Manly, NSW 2095, Australia}
\affiliation[5]{Visiting Scientist at CSIRO Space and Astronomy, PO Box 1130, Bentley, WA 6102, Australia}
\affiliation[6]{CSIRO Space and Astronomy, PO Box 76, Epping NSW 1710 Australia}
\emailAdd{philip.edwards@csiro.au}
\affiliation[7]{National Astronomical Observatory, C. de Alfonso XII, 3, 28014 Madrid, Spain}
\emailAdd{c.garciamiro@oan.es}
\affiliation[8]{INAF Istituto di Radioastronomia, via Gobetti 101, I-40129 Bologna, Italy}
\emailAdd{marcello.giroletti@inaf.it}
\affiliation[9]{Faculty of Aerospace Engineering, Delft University of Technology, Kluyverweg 1, 2629\,HS Delft, The Netherlands}
\emailAdd{leonid@gurvits.org}
\affiliation[10]{Joint Institute for VLBI ERIC, Oude Hoogeveensedijk 4, 7991\,PD Dwingeloo, The Netherlands}
\affiliation[11]{Julius-Maximilians-Universit{\"a}t W{\"u}rzburg, Fakult{\"a}t für Physik und Astronomie, Institut für Theoretische Physik und Astrophysik, Lehrstuhl für Astronomie, Emil-Fischer-Str. 31, D-97074 W{\"u}rzburg, Germany}
\emailAdd{matthias.kadler@astro.uni-wuerzburg.de}
\affiliation[12]{Department of Physics, Ulsan National Institute of Science and Technology (UNIST), 50 UNIST-gil, Eonyang-eup, Ulju-gun, Ulsan 44919, Republic of Korea}
\emailAdd{jaeyoungkim@unist.ac.kr}
\affiliation[13]{Lebedev Physical Institute, Profsoyuznaya 84/32, Moscow 117997, Russia}
\affiliation[14]{Institute of Astrophysics, Foundation for Research and Technology -- Hellas, Voutes, 70013 Heraklion, Greece}
\emailAdd{liodakis@ia.forth.gr}
\affiliation[15]{Shanghai Astronomical Observatory, Chinese Academy of Sciences, 80 Nandan Road, Shanghai 200030, China}
\emailAdd{zshen@shao.ac.cn}
\emailAdd{yuanqi@shao.ac.cn}
\affiliation[16]{Black Hole Initiative, Harvard University, 20 Garden St, Cambridge, MA 02138, USA}
\emailAdd{alexander@plav.in}
\affiliation[17]{Moscow Institute of Physics and Technology, Institutsky per. 9, Dolgoprudny, Moscow region 141700, Russia}
\emailAdd{avpopk@gmail.com}
\affiliation[18]{Crimean Astrophysical Observatory, 298409 Nauchny, Crimea}
\emailAdd{pushkarev.alexander@gmail.com}
\affiliation[19]{Aalto University Department of Electronics and Nanoengineering, PL\,15500, FI--00076 Aalto, Finland}
\affiliation[20]{Aalto University Metsähovi Radio Observatory, Metsähovintie 114, FI--02540 Kylmälä, Finland}
\emailAdd{tuomas.k.savolainen@aalto.fi}
\affiliation[21]{State Key Laboratory of Radio Astronomy and Technology, A20 Datun Road, Chaoyang District, Beijing, 100101, PR China}
\affiliation[22]{Interdisziplin\"ares Zentrum f\"ur Wissenschaftliches Rechnen (IWR), Universit\"at Heidelberg, Im Neuenheimer Feld 205, 69120 Heidelberg, Germany}
\emailAdd{efthalia.traianou@iwr.uni-heidelberg.de}

\abstract{
To date, two dedicated Space Very Long Baseline Interferometry (SVLBI) missions, the VLBI Space Observatory Programme (VSOP) and RadioAstron, have provided groundbreaking insights into the Universe at angular resolutions as fine as $\sim$10 microarcseconds. The phased SKA-Mid, with its exceptional sensitivity and broad frequency coverage, will form a unique ground-based anchor for future SVLBI missions, driving major advances into previously unexplored regions of the angular resolution--sensitivity parameter space.
The discovery of extreme brightness temperatures in blazars by RadioAstron demands detailed investigation with next-generation SVLBI. Such studies are crucial for understanding particle (re-)acceleration mechanisms, with direct implications for the search for high-energy neutrino sources. Combining centimeter-wavelength SVLBI with millimeter ground-based VLBI at comparable resolutions will enable detailed studies of plasma stratification and instabilities in Active Galactic Nuclei (AGN) jets, as well as the processes of jet formation, acceleration, collimation, and magnetic field evolution, for example through Faraday rotation mapping.
The unprecedented sensitivity of the SKA-Mid will allow observations of active galactic nuclei to very high redshifts, tracing their evolution and overcoming opacity caused by the $(1+z)$ shift of intrinsic emission frequencies. Future centimeter SVLBI experiments will also probe scattering in the interstellar medium through pulsar, maser, and AGN observations.
Finally, the combination of multiple tied-array beams from the SKA telescopes and the extremely long SVLBI baselines will enable ultra-precise astrometry using the next-generation MultiView technique, allowing measurements of extragalactic parallaxes of pulsars and megamasers, proper motions of supermassive black holes, and even the astrometric detection of exoplanets.
}

\begin{document}
\maketitle

\section{Introduction}

Breakthroughs in astronomical research are invariably driven by advances in observational capabilities. Scientific progress often operates on a logarithmic scale; consequently, transformative discoveries typically require improvements in one or more key observational parameters by at least an order of magnitude. The SKAO adheres to this ``order-of-magnitude'' principle with respect to sensitivity. However, the integration of the ground-based SKA telescopes with a space-borne radio telescope will establish a Space Very Long Baseline Interferometry (SVLBI) system with unprecedented observational power. 

To achieve high-quality well-sampled ($u$,$v$) coverages, co-observations with other ground-based VLBI telescopes are essential. As a southern-hemisphere facility, the SKAO will yield best observing results for sources at southern declinations. For far southern targets, the common visibility with existing northern-hemisphere ground-VLBI arrays is limited so that the Southern-Hemisphere Long Baseline Array \citep[LBA;][]{Edwards2015} will play an important role in many space-VLBI SKA-Mid experiments. An example of typical ($u,v$) coverages that can be achieved with the current LBA and associated telescopes for different source declinations can be found in \cite{Benke2024}. In the future, these can be substantially improved through the development of a continental African VLBI Network \citep[AVN,][]{Bempong-Manful01.2026.SKA}. At more northern declinations, the European VLBI Network \citep[EVN;][]{Venturi2020} can act as an additional ground-array and future enhancements \citep{Kadler02.2026.SKA} can support synergies with the upcoming next generation Very Large Array \citep[ngVLA;][]{Murphy2018}.
Such a configuration, combining a future SVLBI system with a powerful ground array that includes the SKA telescopes, will extend the accessible parameter space in the sensitivity--angular resolution domain, enabling the exploration of regimes that have thus far remained beyond reach.

\subsection{Previous missions}

Discussions on achieving ultra-high angular resolution by placing at least one element of a radio interferometer in space began concurrently with the early development of Very Long Baseline Interferometry (VLBI) on the ground in the 1960s \citep[][and references therein]{LIG-2023HISTELCON}. The first VLBI “fringes” on baselines exceeding the Earth’s diameter were demonstrated in 1986 by the Orbital VLBI experiment, which utilised NASA’s Tracking and Data Relay Satellite System (TDRSS) geostationary spacecraft in conjunction with a network of ground-based telescopes \citep{Levy+1986Sci}. Observations at 2.3 and 15 GHz, with projected baselines extending up to 2.2 Earth diameters, yielded detection of compact components in about a dozen of quasars. The main achievement of the TDRSS experiment was a confirmation of feasibility of Space VLBI.

In the aftermath of the TDRSS VLBI demonstration, the VLBI Space Observatory Programme (VSOP) was initiated by the Institute of Space and Astronautical Science in Japan in the late 1980s, culminating in the launch of the Highly Advanced Laboratory for Communications and Astronomy (HALCA) satellite in February 1997 \citep{Hirax+1998Sci}. Although nominally a test of the new M-V rocket and the HALCA platform, the mission’s core aim was scientific, pioneering space-based VLBI. HALCA carried an 8.8-metre parabolic radio telescope with gold-coated molybdenum wire mesh and receivers at 18, 6, and 1.3 cm (Left Circular Polarisation, LCP, only). Operating in a medium-eccentricity orbit with a 22,000\,km apogee and 6.3-hour period, the mission achieved an unprecedented data downlink rate of 128~Mbps. Signal coherence was maintained via a hydrogen maser-driven phase-locked loop at ground stations. While the 1.6 and 5~GHz bands functioned effectively, the 22~GHz band suffered irreparable waveguide damage, likely from launch vibrations. Supported by an international consortium (VISC, VSOP International Science Council), VSOP involved a global network of five tracking stations (Japan, Australia, Spain and two in the USA), correlation facilities in Canada, Japan, and the USA, and extensive global Earth-based collaboration. HALCA operated until 2003, yielding sub-milliarcsecond resolution data on compact galactic and extragalactic sources, including hydroxyl spectral lines, thus establishing a foundation for future space VLBI missions.

The RadioAstron mission, a Russian-led successor to VSOP, was launched from Baikonur in July 2011 \citep{NSK+2013AR}. Managed by the Astro Space Center of the Lebedev Physical Institute and the Lavochkin Association, its core instrument was a 10-m parabolic radio telescope, comprising 27 carbon-fibre petals and a central circular section, equipped with dual-polarisation receivers (LCP and RCP) at 0.327, 1.6, 5, and 22~GHz. The satellite operated in a highly eccentric, evolving orbit with an apogee of up to 350,000 km ($\sim$28 Earth diameters) and a 9-day orbital period. Signal coherence was ensured via a phase-locked loop driven by both Earth-based and onboard hydrogen masers, with one spaceborne frequency standard operating reliably for six years. The RadioAstron International Science Council (RISC) coordinated international scientific participation, supported by data acquisition stations in Pushchino (Russia) and Green Bank (West Virginia, USA), and correlation centres in Moscow (Russia), Bonn (Germany) and Dwingeloo (The Netherlands). Following in-orbit validation, the mission adopted an open-access policy, issuing six annual proposal calls. Achieving unprecedented angular resolutions, including a record $\sim$7~$\mu$as at 22~GHz, RadioAstron concluded operations in 2019 after 7.5 years—exceeding its design life by over 2.5 times.

\begin{figure}[t]
\centering
\includegraphics[width=1.0\linewidth, trim=0cm 0cm 0cm 0cm]{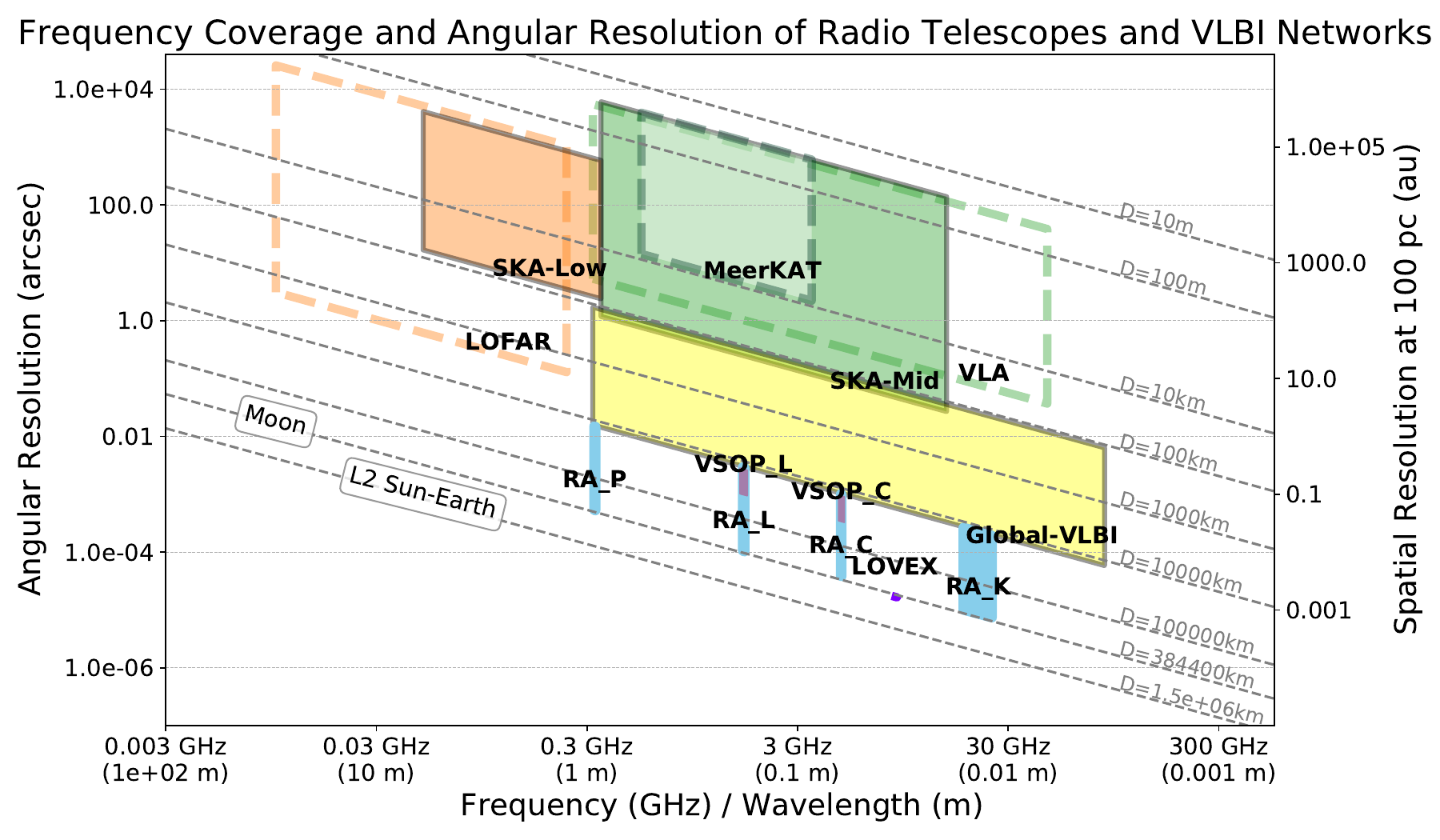}
\caption{The summary figure comparing the frequency coverage and angular resolution by several key ground based and space VLBI instruments.}
\label{fig:SVLBI_resolution}
\end{figure}

The first-generation SVLBI missions, VSOP and RadioAstron, built upon the TDRSS-based demonstration, have established a robust technological foundation for future spaceborne radio interferometry. \autoref{tab:specs} and \autoref{fig:SVLBI_resolution} summarise the principal technical specifications of these three pioneering SVLBI missions.

\begin{table}[h!]
\centering
\caption{Major specifications of SVLBI missions which operated in the period 1986--2019.}
\label{tab:specs}
\begin{tabular}{lccc}
\hline\hline
 & \textbf{TDRSS} & \textbf{VSOP} & \textbf{RadioAstron} \\
\hline
In-orbit operations & 1986--1988 & 1997--2003 & 2011--2019 \\
Antenna diameter [m] & 5.8 & 8.8 & 10 \\
Max baseline $B_\text{max}$ [ED]$^a$ & 2.2 & 2.7 & 28 \\
Operational frequencies [GHz] & 2.3, 15.0 & 1.6, 5.0 & 0.327, 1.6, 5.0, 22 \\
Data rate [Mbps] & 28 & 128 & 128 \\
\hline
\multicolumn{4}{l}{$^a$ Maximum baseline projection on the image plane in Earth Diameters (ED).}
\end{tabular}
\end{table}

At the highest angular resolutions achievable, the three missions listed in \autoref{tab:specs} addressed three principal scientific domains:
\begin{enumerate}
    \item \textbf{Physics of compact continuum sources:} Predominantly extragalactic Active Galactic Nuclei (AGN) powered by supermassive black holes. Though the synchrotron mechanism explains their electromagnetic output, critical physical details remain elusive. Resolving these sub-parsec-scale structures across giga-parsec distances demands microarcsecond angular resolution.
    \item \textbf{Ultra-compact maser sources:} Cosmic masers, notably from H\textsubscript{2}O at 22~GHz and OH at 1.6~GHz, exhibit narrow-band spectral line emission. Earth-based VLBI lacks sufficient resolution to probe the fine-scale morphology of maser emission regions in both galactic and extragalactic contexts \citep[e.g.,][]{2022NatAs...6..976B}.
    \item \textbf{Radio signal propagation studies:} These involve characterising media across cosmic scales—from the interplanetary and interstellar to the intergalactic and near-Earth environments—through analysis of signal distortions.
\end{enumerate}

VSOP's scientific contributions are detailed in~\cite{VSOP-2000aprs} and \cite{VSOP2-2009ASPC}. 
Selected RadioAstron results on AGN, masers, pulsars, and scattering can be found in the following papers: 
\citet{2016ApJ...820L...9K,2018NatAs...2..472G,2023NatAs...7.1359F,2022NatAs...6..976B,2020ApJ...888...57P}.
A series of papers can be found in a volume of Advances in Space Research, dedicated to ``High-resolution space-borne radio astronomy''
\citep{AdSpR-2020}. Collectively, the three missions described above, complemented by advancements in Earth-based radio astronomy, underscore the scientific necessity of future SVLBI systems.

\subsection{Current and future missions}

In the context of the first phase of SKA-Mid, the most relevant development in SVLBI is the implementation of the Chinese Lunar Orbital VLBI Experiment (LOVEX, \citealp{Hong+2025}).
This experiment was deployed aboard the selenocentric relay satellite QueQiao-2 in March 2024. Its primary goal is to support the Chinese Lunar Exploration program. However, the nominal payload of QueQiao-2 was amended with VLBI-specific instrumentation enabling interferometric observations with Earth-based radio telescopes at 8.4~GHz and baseline comparable to the distance Earth--Moon.
LOVEX, together with a network of Chinese radio-telescopes,  
obtained interferometric fringes on Moon--Earth physical baselines in observations of the blazar AO~0235+164.
The experiment also successfully tracked the Chang'E-6 probe across baselines extending up to 33~Earth diameter ($\approx$ 380,000 km) in the near-field VLBI mode -- the first demonstration of this mode on extra-terrestrial baselines. With 512\,MHz maximum bandwidth, an on-board passive H-maser and 4-Terabit onboard data recording systems, LOVEX has validated critical technologies for future space VLBI missions \citep{Hong+2025}.

China's Space VLBI program has evolved into a comprehensive initiative spanning enhanced ground networks, operational lunar-orbit technology demonstrators, and two flagship space-based arrays now in advanced design phases \citep{2020AdSpR..65..850A,2014AcAau.102..217H}, targeting
an angular resolution of 20 microarcseconds. Space Millimetre-wavelength VLBI Array's (SMVA) primary scientific objectives include direct imaging of surroundings of supermassive black hole in M\,87, resolving the innermost jet launch regions in active galactic nuclei \citep{1999Natur.401..891J}, and conducting precision measurements of water megamasers in the accretion disk of NGC\,4258 \citep{2022NatAs...6..976B}.

Another ambitious mission, the Cosmic Microscope (CM), envisions a pair of 30-meter telescopes positioned in extended elliptical orbits (90,000 km $\times$ 2,000 km), operating across a broad frequency range of 30--1670 MHz  \citep{2019ChJSS..39..242A}. This configuration would deliver angular resolutions from 0.4 milliarcseconds at 1.6 GHz to 20 milliarcseconds at 30 MHz, while enabling seamless integration with major ground-based facilities including the Five hundred meter Aperture Spherical Telescope (FAST) and the SKA telescopes. CM's scientific mission focuses on precise positioning of pulsars and radio transients, resolving SMBH binaries, characterizing fast radio burst environments, and detection of magnetic fields from exoplanets. The CM concept overlaps in frequency domain and offers broad scientific synergy with SKA-Low.

We note also missions, concepts or initiatives related to ground-space or space-based initiatives in millimeter domain, which are not related directly to this chapter
\citep[][]{Roelofs+2019AA, 2021PhyU...64..386N, 2021ExA....51..559G, Capella-2023, Johnson+2024SPIE}.


\section{Space VLBI --- SKAO science cases}

\subsection{High-energy jet physics, particle acceleration, and multi-messenger connections}

\paragraph{Localizing the sites of high--energy neutrino emission}

Recently, VLBI images of parsec-scale jets have helped identify highly
beamed blazars as the most likely counterparts to IceCube events \citep{2025ApJ...991...33P,2025A&A...700L..12K}.
They also show tentative time correlations between neutrino detections and radio flares \citep{2024A&A...690A.111K}, possibly parsec-scale ejections of components.
A future SKA-SVLBI array would build on this approach and could,
for the first time, provide the sensitivity and resolution needed to test
these suggested links within a fully \emph{multi-messenger} framework.  
With one pointing taken a few days after an alert, it would cover the
entire square-degree localization region while resolving every candidate
source on ${\sim}\,0.1$\,mas (sub-parsec) scales.  
Microjansky sensitivity would reveal newborn, still partially self-absorbed knots at
or very near the moment of their launch, much closer to the jet origin than
ground arrays can reach.  
Monitoring the structural and flux density evolution of all candidates
simultaneously should establish clear sub-parsec links between neutrino
production and the birth of individual radio components.

\paragraph{Diagnosing particle--acceleration mechanisms in AGN jets}
Magneto-hydrodynamic (MHD) models predict two key transitions on sub-parsec
scales.
First, as the jet widens from parabolic to conical, magnetic fields
no longer drive bulk acceleration and the Lorentz factor stops growing.
Second, the energy budget changes from magnetically dominated
($\sigma \gg 1$, 
where $\sigma$ is the magnetization parameter --- see \autoref{s:RM}) to particle dominated ($\sigma \lesssim 1$).
As magnetic energy is dissipated, the resulting change in the energy budget can accelerate non-thermal particles that produce the observed synchrotron emission.

Ground-based VLBI has already detected the parabolic-to-conical break in a few nearby radio galaxies. Applying the same test to powerful, distant blazars will need higher sensitivity and finer resolution. With an ${\sim}0.1$\,mas beam a future SKA-SVLBI array could follow the jet radius $R(d)$ as a function of de-projected distance $d$, identifying the change from $R\!\propto\! d^{0.5}$ to $R\!\propto\! d^{1}$ even at cosmological distances. 
Imaging the partially opaque cores at multiple frequencies would allow synchrotron turnover mapping, from which the magnetic field $B(d)$ and particle density $n(d)$ can be estimated.

Measuring both transitions in the same objects would show how energy is first converted into bulk motion and later shared with the energetic particle population that shapes the high-frequency spectrum, and that may also contribute to the cosmic-ray and neutrino output of blazars.

\paragraph{Resolving the brightness--temperature crisis}

Earlier space VLBI missions have reported extremely high brightness temperatures reaching $10^{13}$--$10^{14}$\,K, values which physical origin remains unclear \citep[e.g.,][]{2000PASJ...52..975F,2016ApJ...820L...9K}.
Whether they reflect extreme Doppler boosting, proton–synchrotron emission
or coherent plasma processes cannot be decided without greater sensitivity
and dynamic range at the highest angular resolutions.  
A future SKA-SVLBI array, providing ${\sim}\,0.1$\,mas resolution
combined with microjansky noise levels, would measure $T_\mathrm{b}$ to
within $\pm 0.1$\,dex while simultaneously imaging the surrounding jet
structure.  
Placing the high-brightness knots within their wider morphological and
kinematic context is essential for understanding how, where and under what
conditions such extreme emission arises, and for assessing its role in the
multi-messenger life cycle of high-energy particles.

\subsection{Faraday rotation measure and magnetic field}
\label{s:RM}

Faraday rotation measurements offer a direct probe of magnetic field topology in AGN jets. Multi-frequency VLBI observations have revealed systematic transverse RM gradients indicating organized helical fields \citep{OSullivan2009,Hovatta2012}, with rotation measures from hundreds to tens of thousands of rad\,m$^{-2}$ concentrated near jet boundaries \citep{Kravchenko2017}. These patterns point to toroidal field components that may be generated by disk rotation, black hole spin, or current sheets within the jet if the Faraday rotation is internal.
Yet extracting this information has proven difficult: robust RM synthesis demands dense frequency sampling \citep{Brentjens2005}, which remains challenging at $\mu$as scales. SKA-Mid changes this landscape. Its broad bandwidth (0.35--15.4\,GHz) combined with space VLBI resolution enables reliable RM measurements of faint features while resolving the transverse structure where toroidal signatures peak. With this capability, we can directly test MHD models predicting how the toroidal-to-poloidal field ratio varies with distance from the central engine.
Going further, combining RM with core-shift or synchrotron-turnover measurements largely breaks the degeneracy between magnetic field strength and particle density, yielding field strengths of the order of $0.1$-$1$~G on parsec scales.
To translate these into a magnetization we must assume a plasma content: for a mixed pair-proton jet (rather than a purely leptonic flow), the resulting magnetization parameter \(\sigma = B^2 / (8\pi n m_p c^2 \Gamma^2)\) (where \(B\) is the magnetic field strength, \(n\) is the particle number density, \(m_p\) is the proton rest mass, \(c\) is the speed of light, and \(\Gamma\) is the bulk Lorentz factor) identifies where the flow changes from magnetically dominated (\(\sigma \gg 1\)) to particle dominated (\(\sigma \lesssim 1\)).
Different, but still plausible, compositions (pair-rich, heavy-ion) would shift the absolute $\sigma$ values but not the ability of SKA-SVLBI to map the radial trend of magnetization.

Beyond static snapshots, time-variable Faraday rotation reveals jet dynamics directly. Multi-epoch observations show RM variations on week-to-year timescales, often correlating with high-energy flares \citep{Hovatta2012}, with propagating shocks creating favorable conditions for particle acceleration \citep[e.g.,][]{2010ApJ...715..362J, 2020A&A...634A.112T}. Systematic SKA-SVLBI monitoring will distinguish jet-intrinsic variations from external Faraday screens, potentially establishing whether extreme RM values ($>10^5$\,rad\,m$^{-2}$) trace magnetically dominated outflows near supermassive black holes.

Ground-based and space VLBI experiments have provided significant evidence for the magneto-hydrodynamic launch and high magnetization of AGN relativistic jets at their base \citep{2024ApJ...973..100K, 2025A&A...693A...9L}.
Millimeter and sub-millimeter polarimetric studies enable the nature and source of the magnetized media to be derived through the rotation measure (RM) and wavelength-dependency of the polarization emission. 
Provided the high angular resolution of the VLBI observations  resolves the depolarization issues, the orientation of the magnetic field in the inner jet regions can be revealed.
These quantities can also be used to probe accretion flows and winds around black holes \citep{2024A&A...682A.154T,2024A&A...682L...3P,2019ApJ...871..257P}.
The mass loss via winds could have a substantial influence on the measured accretion rate compared to the actual rate of mass accreted onto the black hole.

\subsection{Plasma stratification and instabilities in AGN jets: a combination of centimeter SVLBI and millimeter ground VLBI arrays at similar angular resolutions}

Another long-standing open question is the nature of jet components and the transverse jet structure. High-resolution high-dynamic-range EHT \citep{2020A&A...640A..69K} and RadioAstron \citep{2023NatAs...7.1359F} observations showed filamentary structure of blazar jets in contrast to the uniform distribution of emission across the jet seen in ground-based VLBI studies.
Centimeter SVLBI and ground-based millimeter VLBI can achieve comparable angular resolutions (tens of $\mu$as), enabling direct, frequency-dependent tomography of AGN jets from a few to $\sim10^4$ Schwarzschild radii. This synergy exposes transverse plasma stratification (spine--sheath), jet collimation/acceleration profiles, and the imprint of current-driven and Kelvin–Helmholtz (KH) instabilities in unprecedented detail \citep{2018NatAs...2..472G,Boccardi2015,2023A&A...676A.114S}.

\paragraph{Matched-resolution views of the jet base.} 
Disentangling these structures requires the matched-resolution approach that centimeter SVLBI and ground-based millimeter VLBI together make possible.
At 86\,GHz, GMVA images of M\,87 resolve the limb-brightened jet down to $\sim7\,R_{\rm S}$, with a compact core of $\sim(8$--$13)\,R_{\rm S}$ and brightness temperature $T_{\rm b}\sim(1$--$3)\times10^{10}$\,K, implying magnetic-energy dominance at launch and a sheath width of order a few $R_{\rm S}$ \citep{2018A&A...616A.188K,Lu23}. SVLBI at cm wavelengths complements these views: RadioAstron baselines ($\lesssim$\,few–8 Earth diameters) delivered $\sim$30\,$\mu$as resolution at 22\,GHz, resolving a limb-brightened flow in 3C\,84 down to $\sim$175\,$R_{\rm S}$ and revealing a very broad jet ($\gtrsim125\,R_{\rm S}$) that either expands rapidly within $<50\,R_{\rm S}$ or originates in the disk-launched sheath \citep{2018NatAs...2..472G}. In blazars, RadioAstron imaging of 3C\,279 uncovered multiple narrow filaments rooted at the core, consistent with a helical field and a kinetically dominated, unstable flow with $\Gamma\sim13$ and an intrinsic pitch angle of $\sim45^\circ$ \citep{2023NatAs...7.1359F}.

Furthermore, SVLBI at cm wavelengths will provide a unique view of the true nature of nuclei of low-luminosity AGNs (LLAGNs).
LLAGNs (and even $\sim1/3$ of centers of bright nearby galaxies; \citealt{heckman80}), can contain compact radio nuclei without signatures of strong jets \citep{ho08}.
Whether the physical origin of the compact radio emission is due to weak outflow activity, advection dominated accretion flow, or star formation has long been debated. 
While multi-frequency VLBI analysis such as the core-shift measurement and frequency-dependent core size measurement can help distinguish its origin (e.g., \citealt{yan24}), those analyses significantly suffer from systematic errors in the core-jet discrimination and core size measurement, due to the poor resolution at the long-wavelength end. 
With the superb sensitivity of SKA-Mid and extreme angular resolution to be achieved by the SVLBI, multi-frequency and high-resolution observations of LLAGNs will become possible at matched resolution, allowing us to identify the true nature of compact radio nuclei in a vast majority of LLAGNs. 

\paragraph{Transverse stratification and frequency dependence.} SVLBI of 3C\,273 shows limb-brightened edges at 1.6\,GHz, but a centrally peaked stream at 4.8\,GHz; the complementary structures demand stratification in the emitting electron distribution (and/or velocity/magnetic structure) across the jet \citep{2021A&A...654A..27B}. At 7\,mm, Cygnus\,A exhibits a two-sided, limb-brightened flow with distinct acceleration gradients for a slow sheath and faster spine, consistent with magnetically driven jets in near-equilibrium collimation \citep{Boccardi2015}. Together, these results show that matched-resolution cm--mm imaging isolates the sheath at lower frequencies (where optical depth and Doppler selection favor edges) and the spine at higher frequencies, enabling a reconstruction of jet layering.

\paragraph{Collimation and acceleration.} 
Parabolic-to-conical transitions, previously studied in low-power radio galaxies are now established in powerful systems: in M\,87, the jet narrows with a parabolic profile inside the Bondi scale before becoming conical, indicating externally confined, magnetically accelerated flow that relaxes downstream \citep{Asada2012}. In Cygnus\,A, the acceleration region extends to $\sim10^4\,R_{\rm S}$ with $r\propto d^{0.55\pm0.07}$ (where $d$ is the de-projected distance), as expected for MHD jets in pressure equilibrium with an external medium ($p\propto d^{-k},\,k\lesssim2$) \citep{Boccardi2015}.
This nicely matches the standard model of jet formation, collimation and acceleration. In this model, hot, geometrically thick accretion flows around rapidly spinning black holes launch magnetized outflows that are subsequently accelerated by magnetic pressure gradient and (self-)collimated by magnetic hoop stress and pressure from their surrounding medium \citep[see also a sample study by][]{2020MNRAS.495.3576K}.
In such systems, typically giant ellipticals, the interplay between magnetic tension and external confinement leads to gradual acceleration and a transition from magnetically dominated to kinetically dominated flow (see \citealt{blandford19}). However, recent theoretical studies involving advanced General Relativistic Magnetohydrodynamics (GRMHD) simulations have shown that jet formation may also occur in a broader range of accretion and magnetic field geometries, including thin disks with toroidal or poloidal magnetic flux (e.g., \citealt{liska20,dihingia21,lalakos24}; see also \citealt{mizuno22}).
They motivate a systematic, high-resolution observational study of many black holes, resolving spatial scales close to the jet-launching region and within the sphere of gravitational influence, in order to establish a comprehensive understanding of the jet formation scenario across diverse accretion environments.

The SKA-SVLBI will uniquely enable detailed studies of the collimation and acceleration of jets in rare and diverse systems through its combination of extreme sensitivity and sub-milliarcsecond resolution, allowing high-fidelity imaging of faint and distant jets, including those in spirals, Seyferts, and radio-quiet quasars where only weak or compact outflows have been detected so far. 
The SKA-Mid alone will discover a large population of new compact radio sources and produce extensive radio-optical matched catalogs, suitable for detailed high-resolution follow-up. These discoveries will open the way for systematic studies of jet formation across a much broader range of host types, luminosities, and accretion states.
Crucially, SKA-SVLBI--level angular resolution will be required to probe down to scales comparable to or within the Bondi radius for a significant number of nearby systems, a regime that will remain challenging for ground-based SKA--VLBI alone or for higher-frequency ground VLBI experiments due to the steep synchrotron spectra of the sources. 

\paragraph{Instabilities and magnetic topology.} SVLBI observations of 3C\,273 reveal a \emph{double-helix} internal pattern that matches multi-mode KH instability in a light, mildly relativistic jet, direct evidence that shear-driven modes can organize jet substructure on parsec scales \citep{Lobanov2001}. Transverse RM gradients and polarization in 3C\,273 indicate a helical magnetic field, consistent with disk winding and with the ordered fields inferred from filamentary structures \citep{2021A&A...654A..27B}. Such instabilities, coupled with a helical field, plausibly drive the observed filaments and limb-brightening, and can modulate radio variability via differential Doppler boosting along strands rather than solely by shock-in-jet scenarios. Moreover, recent RadioAstron observations have revealed a striking, ribbon-like jet morphology in the BL Lac object OJ\,287, resolved down to unprecedented angular scales of $\sim$15\,\textmu as \citep{2022ApJ...924..122G,2025A&A...700A..16T}. This sinuous jet shows multiple sharp bends within the innermost milliarcsecond, forming a complex trajectory that deviates markedly from a simpler conical flow. The ribbon-like shape can be interpreted as the manifestation of helical KH or current-driven instabilities developing in the spine-sheath jet structure, with the jet’s precession enhancing perturbation growth.

\paragraph{Outlook.} From combined cm SVLBI and mm VLBI observations at matched resolution, the following picture has started to emerge: magnetically dominated jets emerge with spine--sheath stratification, collimate parabolically while accelerating over $\sim10^3$--$10^4\,R_{\rm S}$, and develop KH/Current-Driven signatures—filaments, helices, and limb-brightening, the visibility of which depends on frequency and viewing angle. This joint strategy thus provides multi-band constraints on magnetization, external pressure, and instability growth, linking launch physics to parsec-scale dissipation.

\begin{figure}[t]
\centering
\includegraphics[width=15cm]{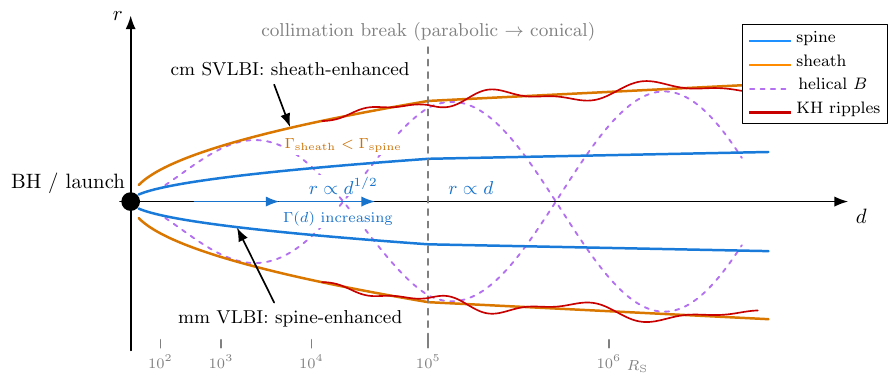}
\caption{Jet launching and collimation: the outflow expands parabolically ($r\!\propto\!d^{1/2}$, where $d$ is the de-projected distance) within the acceleration and collimation zone, then transitions at the collimation break to a conical, freely expanding structure ($r\!\propto\!d$). Small KH ripples mark instability onset near the sheath boundary, with helical magnetic field lines indicating the ordered field structure due to winding from the disk or the black hole ergosphere. The bottom ruler indicates distances in Schwarzschild radii ($R_{\rm S}$). Centimeter SVLBI preferentially traces the sheath, while millimeter ground-VLBI isolates the spine.}
\label{fig:acz_parab_to_cone_final_nopressure}
\end{figure}

\subsection{Galactic nuclei to very high redshift}

AGNs are the most powerful ``engines'' in the Universe, distributed over a range of redshifts reaching at least redshift $z \simeq 11$. Their detailed studies are crucial for understanding the evolution of objects on the galactic scale and the Universe as a whole. 
Investigation of redshift-dependent evolution of compact (milli- and sub-milliarcsecond-scale) structures in AGN requires matching both the angular resolution and emitting frequencies at a broad range of redshift. However, achieving matching angular resolution at a given emitting frequency for low- and high-redshift sources necessitates observations of the latter at lower observing frequencies. In turn, this latter requirement results in deteriorating angular resolution. It becomes a fundamental limitation once the ``matching'' baseline for lower-frequency VLBI observations require baselines exceeding the Earth's diameter. This limitation can be overcome only with baselines exceeding the Earth diameter, i.e., by means of space VLBI. Moreover, as demonstrated by \citet{Spingola01.2026.SKA}, such observations of AGN matched in the emitting frequency and angular resolution at a wide range of redshifts requires enhanced sensitivity of VLBI observations at lower frequencies, thus involving space radio telescopes \citep{LIG-2000pras}. Such a requirement makes the involvement in these observations an Earth-based radio telescopes with ultra-high sensitivity a must. This is precisely the task for the SVLBI with SKA-Mid.

\subsection{Scattering: interstellar medium probed by observations of background sources}

Interstellar scattering is an important consideration for SVLBI observations as it affects the apparent structure, brightness and position of sources on the (sub-)mas scales of interest. Moreover, the sensitivity of an SKA-SVLBI array would enable robust determinations of intrinsic source properties at extremely high angular resolution through measurements of refractive substructure, as well as providing unique probes of interstellar and even intergalactic scattering. 

Diffractive-dominated scattering leading to angular broadening, especially towards lower frequencies for compact sources seen through the Galactic plane \citep{Pushkarev15}, will effectively probe the physical properties of the highly turbulent interstellar medium (ISM). The combination of high-sensitivity and low-frequency SKAO observations will enable a significant increase in the population of scattered sources and thus a modernization of the sky distribution of the scattering power \citep{Koryukova22} at a higher level of accuracy. These low-frequency observations are also crucial to constrain confined scattering structures which crucially affect the frequency-dependence of the angular broadening measurements \citep{Cordes:2001}. A strong candidate to explore this could be Sagittarius A*, which has a strongly anisotropic scattering structure and whose effect on SVLBI observations has been studied by several authors \citep{Johnson:2021,Issaoun21,Tamar:2025}. Herein, the deviations from the commonly assumed frequency-dependent scaling of the angular broadening has been posited \citep{Krichbaum:1998,Lo:1998} at higher frequencies than that of the SKA telescopes. Even at the SKAO frequencies, deviations from the commonly-assumed Kolmogorov-scaling of the frequency dependence has been observed for pulsars at the low Galactic latitude \citep{Bhat:2004}. Thus, the proposed first-generation receiver bands for the SKAO as the ground station for a space VLBI array provides a hitherto unexplored observing framework to explore scattering structures to improve our understanding of effects of interstellar scattering on radio sources.

\citet{2016ApJ...820L..10J} analysed the effect of refractive substructure on RadioAstron detections of 3C\,273, showing that refractive noise due to interstellar scattering at high galactic latitudes has a significant effect on visibilities for SVLBI baselines across the SKA-Mid frequency bands. SKA-SVLBI would enable this effect to be measured for a large number of sources. Measurements at several epochs/frequencies would allow a robust estimate of intrinsic source brightness temperature as well as ISM scattering measures.

Extreme Scattering Events (ESEs), due to refractive lensing of background sources by AU-scale structures in the ISM, are expected to cause multiple imaging and refractive shifts, but to date such measurements have been limited due to the challenges of obtaining high angular resolution observations of ESEs in progress \citep{2013A&A...555A..80P, 2016Sci...351..354B,Koryukova23}. This has severely limited our understanding of ESE lenses. Broadband SKAO surveys are expected to identify many ESEs through spectral variations, and the ultra-precise astrometry enabled by SKA-SVLBI with the MultiView techniques discussed below in \autoref{sec:astrometry} would allow the geometry and angular scale of the lens to be measured.

Although the present chapter focuses primarily on SKA-Mid, SKA-Low also offers unique opportunities for Space VLBI at low frequencies ($\sim$50--350\,MHz). Thanks to its unprecedented sensitivity and capability to form multiple tied-array beams, SKA-Low can serve as a highly sensitive anchor for SVLBI observations of pulsars, OH masers, fast radio bursts (FRBs), and other transient sources, enabling detailed measurements of diffractive and refractive scattering in the interstellar (ISM) and intergalactic (IGM) media \citep{2015MNRAS.446.2370K}. A particularly promising synergy exists with planned missions such as the Cosmic Microscope (CM; \citealt{2019ChJSS..39..242A}), whose frequency coverage below 1.6\,GHz would be naturally complemented by SKA-Low, enabling tomographic studies of scattering screens on scales from parsecs to kiloparsecs. Stronger ionospheric and tropospheric effects and increased source confusion due to the large FoV can be effectively mitigated by the MultiView and phase-referencing techniques discussed in Sect.~\ref{sec:astrometry}. In summary, SKA-Low will extend the SVLBI parameter space towards the lowest radio frequencies, opening windows highly complementary to those accessible with SKA-Mid, with further overlap expected from prospective spaceborne and Moon-based low-frequency facilities \citep[e.g.,][and references therein]{Ghosh+2026MNRAS, wen2026-fs}.

\subsection{The next generation ultra precise astrometry}
\label{sec:astrometry}

Astronomical studies in Space VLBI are particularly challenging, as one of the antennas is literally free-floating in space. A precise location for the antenna is normally a prerequisite for astrometry. 
Furthermore, whilst the error contributions from the atmosphere are reduced, if one end of the baseline is Earth bound atmospheric calibration is still required.
Nevertheless, astrometric studies were performed with the VSOP mission \citep{rioja_astrometry_vsop}, using a pair of sources sufficiently close such that they both fell in the same field of view (FoV) for the space antenna. 
These observations provided an upper estimate for the orbit error, rather than a true astrometrical measurement. 
Any conventional astrometrical observation between a pair of sources will suffer from the same limitations.

The next generation methods for astrometry can fully resolve this issue, by canceling out the contributions from the antenna positions to measurement accuracy. 
Two main classes of methods have been developed, the first more suitable for high frequencies~--- Source/Frequency Phase Referencing (SFPR), based on Frequency Phase Transfer (FPT) approaches \citep{2009arXiv0910.1159D}~--- and the second, MultiView (MV; \citealt{dodson_2013}), which is mainly aimed at SKA-Mid applications but also suitable for all frequency ranges.

The science targets for astrometry in Space-VLBI are a subset of those already discussed in this chapter. 
Astrometric registration provides the ability to compare images between instruments and frequencies; arguably without astrometric registration no rotation measure nor spectral index measurement is robust.
Thus astrometry is vital if one wishes track the evolution of an object as a function of frequency or time; that is to extract insights into the physical conditions or measure a distance.

Several simulation studies have been performed to address the issues for astrometry in space VLBI. 
A crucial one was prepared for the VSOP-2 Mission \citep{asaki_2007}, 
where a conventional phase referencing capability was included in the mission for frequencies as high as 43\,GHz. 
That is, the ability to perform rapid source switching and a very precise orbit reconstruction were designed in. 
The source switching was challenging as it required the angular rotation of a massive free floating device in a very controlled and repeatable manner.
The orbit reconstruction was challenging, in that it required cm-level precision (compared to the several meters of VSOP) for orbits that lay above the GPS constellations. 

However, the next generation methods ameliorate these constraints significantly, as discussed in \citet{rioja_2020}. 
For FPT methods non-dispersive errors, such as positions, are compensated through referencing the high frequency observations against the low frequency observations.
This removes the ability to provide relative spatial astrometry (as the reference is the target itself), but for investigations of physical conditions of the source --- such as rotation measures or astro-chemical environments --- it is more than sufficient. 
Simultaneous frequency observations are now standard practice on instruments such as the Korean VLBI Network. 
\citet{rioja_2011b} discussed SFPR and FPT calibration for astrometry for a VSOP-2--like mission and found 1$\mu$as precision in many configurations.

For MV methods both non-dispersive and dispersive errors are corrected for, and spatial astrometric registration is retained (although to a virtual reference point that is a combination of multiple calibrators). 
MV does require multiple calibrator sources to reference against; three plus the target is sufficient to fully solve for a phase plane across the target field. 
MV is suitable for all frequencies, and corrects for ionospheric, tropospheric and positional errors. 
It does require observations of more reference sources than just the target itself (unlike FPT), but given the small dish size for a space mission there are typically many sources in the FoV. 
The only consideration is the dynamic range that is required to achieve the science goals.
Estimates for 21~cm observations with AA* or AA4 predict that about 4 sources, with a dynamic range greater than 100, would be detectable in a random FoV.
On the other hand, for 6~cm observations there would 
rarely be even one source in a random FoV. 

\begin{figure}[t]
    \centering   \floatbox[{\capbeside\thisfloatsetup{floatwidth=sidefil,capbesideposition={right,top},capbesidewidth=.40\linewidth}}]{figure}
    {\caption{
    Astrometric precision achievable with Space VLBI and MultiView, with two satellites in a Low Earth Orbit, following \citet[Figure 8]{dodson_2013} at L-band.
    The x-axis shows the minimum separation of the target from the MultiView calibration sources, and the y-axis the resultant astrometric error determined from the simulations.
    For 8~m orbit errors, the cyan stars plot the recovered precision for moderate ionospheric conditions and Traveling Ionospheric Disturbances (TID).
    For 8~cm orbit errors, green squares are for a strong Kolmogorov ionospheric disturbance. Red diamonds are for the moderate TID. 
    Plotted with a dotted line are the typical current best performance from in-beam phase referencing.
    \label{fig:astrometry_mv}}}{\includegraphics[width=0.6\textwidth,trim=0.6cm 0cm 1cm 0cm]{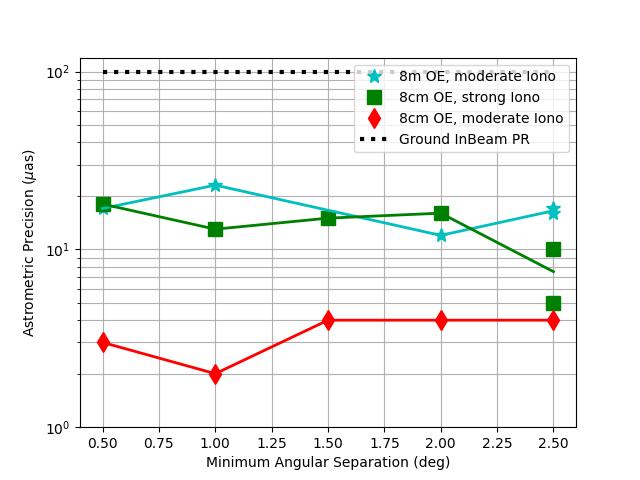}}
\end{figure}

\citet{dodson_2013} discussed MV calibration for astrometry for a twin spacecraft mission (such as Cosmic Microscope) at SKA-Mid frequencies.
\autoref{fig:astrometry_mv}, from \citet{dodson_2013}, shows the astronomical precision that could be achieved, deduced from simulations of MV observations at 1.4\,GHz in a turbulent and realistic atmosphere with ionospheric and tropospheric contributions. High Signal-to-Noise Ratio (SNR) observations were assumed, so these observations have no thermal limits, only the systematic ones from the calibration strategy.
To provide a large FoV the simulations assumed a Phased Array Feed (PAF; based on the ASKAP design) would be used, that directly increased the FoV by a factor of 6 in both directions.
A PAF allows for the formation of multiple beams on the sky and removes the requirement for the calibrators to be in the single-dish FoV, and as such is optimal for the MV method.

The microarcsecond-level astrometric precision achievable with SKA-SVLBI will open new possibilities for detecting and characterizing exoplanets via the reflex motion of nearby stars. Combined with the high sensitivity of SKA-Mid, such observations could complement optical missions like Gaia, extending astrometric planet detection into the radio domain and enabling studies of magnetically active or radio-loud host stars. In a synergy with future space VLBI projects, SKA-Low and SKA-Mid could also probe planetary magnetic fields via directly imaging coherent radio bursts from exoplanetary magnetospheres.

\section{Summary}

At a fixed wavelength, the only way to improve the angular resolution is to observe on longer baselines. 
While flat-spectrum continuum sources may be observed at higher frequencies, that is not feasible for spectral line sources.
The measureable maximum brightness temperature also depends on baseline length, and so the best way to constrain this key parameter is to extend VLBI baselines beyond the Earth’s surface with an orbiting telescope.

Space VLBI telescopes are, by necessity, limited in size, and so require the most sensitive ground telescopes to co-observe with them.
The VSOP and RadioAstron missions were successful through co-observing with ground arrays including the phased Very Large Array (VLA), Arecibo, the Green Bank Telescope (GBT), Effelsberg, and other large apertures, although smaller diameter dishes remain important to obtain good ($u$,$v$) coverage for imaging experiments.
In the SKA era, it is clear that future SVLBI missions will be enhanced by co-observing with ground arrays that include the SKA telescopes.

A number of science cases have been laid out in this chapter, 
including extreme brightness temperatures in blazar cores; the study of AGN jet formation, acceleration, collimation; tracing the evolution of AGN to higher redshifts; probing the scattering properties of the interstellar medium; and potentially achieving ultra-precise astrometry for proper motions and parallaxes.
The SKA telescopes are being built with the capability to participate in VLBI observations, and no additional effort is required to co-observe with SVLBI observations. 

%

\bibliographystyle{abbrvnat-maxbibnames4}
\setlength{\bibsep}{0.0pt} 
\newcommand{\actaa}{Acta Astron.} 
\newcommand{\araa}{ARA\&A} 
\newcommand{\aar}{A\&ARv} 
\newcommand{\aapr}{A\&ARv} 
\newcommand{\ab}{Astrobiol.} 
\newcommand{\aj}{AJ} 
\newcommand{\apj}{ApJ} 
\newcommand{\apjl}{ApJL} 
\newcommand{\apjs}{ApJSS} 
\newcommand{\ao}{Appl. Opt.} 
\newcommand{\apss}{Astro. \& Space Sci.} 
\newcommand{\aap}{A\&A} 
\newcommand{\aaps}{A\&AS.} 
\newcommand{\baas}{Bull. Am. Astron. Soc.} 
\newcommand{\caa}{Chinese A\&A} 
\newcommand{\cjaa}{Chinese J. A\&A} 
\newcommand{\cqg}{Class. Quantum Gravity} 
\newcommand{\gal}{Galaxies} 
\newcommand{\gca}{Geo. Cosmo. Acta} 
\newcommand{\icarus}{Icarus} 
\newcommand{\jcap}{JCAP} 
\newcommand{\jgr}{J. Geophys. Res.} 
\newcommand{\jgrp}{J. Geophys. Res. Planets} 
\newcommand{\jqsrt}{J. Quant. Spectrosc. Radiat. Transf.} 
\newcommand{\memsai}{Mem. SAIt} 
\newcommand{\mnras}{MNRAS} 
\newcommand{\nat}{Nature} 
\newcommand{\nastro}{Nat. Astron.} 
\newcommand{\ncomms}{Nat. Commun.} 
\newcommand{\nphys}{Nat. Phys.} 
\newcommand{\na}{New Astron.} 
\newcommand{\nar}{New Astron. Rev.} 
\newcommand{\physrep}{Phys. Rep.} 
\newcommand{\pra}{Phys. Rev. A} 
\newcommand{\prb}{Phys. Rev. B} 
\newcommand{\prc}{Phys. Rev. C} 
\newcommand{\prd}{Phys. Rev. D} 
\newcommand{\pre}{Phys. Rev. E} 
\newcommand{\prx}{Phys. Rev. X} 
\newcommand{\prl}{Phys. Rev. Let.} 
\newcommand{\psj}{Planet. Sci. J.} 
\newcommand{\planss}{Planet. Space Sci.} 
\newcommand{\pnas}{Proc. Natl Acad. Sci. USA} 
\newcommand{\procspie}{Proc. SPIE} 
\newcommand{\pasa}{PASA} 
\newcommand{\pasj}{PASJ} 
\newcommand{\pasp}{PASP} 
\newcommand{\rmxaa}{RMXAA} 
\newcommand{\sci}{Science} 
\newcommand{\sciadv}{Sci. Adv.} 
\newcommand{\solphys}{Sol. Phys.} 
\newcommand{\sovast}{Soviet Ast.} 
\newcommand{\ssr}{Space Sci. Rev.} 
\newcommand{\uni}{Universe} 

\bibliography{svlbi}
\end{document}